\documentclass{ws-ijbc}

\usepackage{graphicx}
\usepackage{amssymb}
\usepackage{color}
\definecolor{red}{rgb}{0.793,0.238,0.336}
\definecolor{black}{rgb}{0,0,0}

 \usepackage{lineno}

\begin{document}
\catchline{}{}{}{}{} 

\title{
How to avoid potential pitfalls in recurrence
plot based data analysis}
\author{Norbert Marwan}

\address{Potsdam Institute for Climate Impact Research,
P.O.~Box 601203, 14412 Potsdam, and\\
Interdisciplinary Centre for Dynamics of Complex Systems, 
University of Potsdam, 14415 Potsdam\\ Germany\\
marwan@pik-potsdam.de}
\maketitle

\begin{history}
\received{(to be inserted by publisher)}
\end{history}

\begin{abstract}
Recurrence plots and recurrence quantification analysis
have become popular in the last two decades. Recurrence
based methods have on the one hand a deep foundation in 
the theory of dynamical systems and are on the other hand
powerful tools for the investigation of a variety of problems.
The increasing interest encompasses the growing risk of 
misuse and uncritical application of these methods. Therefore,
we point out potential problems and pitfalls related to
different aspects of the application of recurrence plots
and recurrence quantification analysis.
\end{abstract}

\keywords{recurrence plot, recurrence quantification analysis, time series analysis, pitfalls}

\section{Introduction}

Since its introduction in 1987 by \citet{eckmann87}, and the development
of different quantification approaches, recurrence plots (RPs) have been
widely used for the investigation of complex systems in a variety of different
disciplines, as physiology, ecology, finance or earth sciences
\cite[e.g.,][]{marwan2008epjst,schinkel2007,zbilut2004b,facchini2007,belaire2004,pecar2003,trauth2003,cermak2009}. 
RPs may attract attention because of their ability to produce
beautiful or fancy pictures, as in the case of the colourful
representations of fractal sets \cite{mandelbrot82}.
The recent remarkable increase of applications can be traced 
down in part to several free software packages available for calculating
recurrence plots and the corresponding recurrence quantification 
analysis (RQA). Since these methods are also
claimed to be very powerful even for short and non-stationary 
data, we should be careful not to consider them as a kind of a magic tool,
which works on all kinds of data. Owing to the fact that 
these methods are indeed in some sense powerful and rather 
adaptable to various problems, it is really important that the
user knows how these methods work and has understood the ideas 
behind the RP and the measures of complexity derived from it.
Any uncritical application
will lead to serious pitfalls and mis-interpretations.
As the number of applicants increases, the risk of careless 
application of RPs and RQA grows. 

In this article we try to highlight some of the pitfalls which can
occur during the application of RPs and RQA and present future
directions of research for a deep theoretical understanding
of the method.

\section{Recurrence plots and recurrence quantification}
Although similar methods already existed before, the RP,
$R_{i,j} = \Theta(\varepsilon - ||\vec x_i - \vec x_j||)$, 
for the analysis of the dynamics of a 
dynamical system by using its phase 
space trajectory was introduced by \citet{eckmann87}. 
This method can be used
in order to visualise the recurrence of a state, i.e., all the
times when this state will recur. In the 
1990's, a heuristic approach of
quantification RPs by its line structures
has led to the recurrence quantification analysis (RQA) 
\cite{webber94,marwan2002herz}. In this approach, the 
density of recurrence points as well as the histograms $P(l)$ of
the lengths $l$ of the diagonal and vertical lines in the RP are quantified.
The density of recurrence points (recurrence rate) coincides
with the definition of the correlation dimension \cite{grassberger83c}.
Moreover, RPs contain much more information about the dynamics of 
the systems: dynamical invariants like R\'enyi entropy or 
correlation dimensions can be derived from the structures in
RPs \citep{faure98,thiel2004a}, RPs can be used to study synchronisation
\cite{romano2005,senthilkumar2006} or to construct surrogate
time series \cite{thiel2008} and long time series from ensemble 
measurements \cite{komalapriya2010}. For a comprehensive introduction
we point to \cite{marwan2007}.

\section{Pitfalls}

\subsection{Parameter choice for recurrence analysis}

RP and RQA depend
on some parameters which should be properly chosen. For the
actual recurrence analysis, a recurrence threshold is necessary.
This measure is probably the most crucial one and is discussed
in the next subsection.

As already mentioned, the quantification of recurrence structures depends on lines
in the RP; by defining a minimal length of such lines, it
is possible to adjust the sensitivity of line based recurrence
measures. In Subsect.~\ref{sec_indic_det} and \ref{sec_indic_period}
we will come back to this parameter.

\begin{figure}[bth]
\centering \includegraphics[width=\columnwidth]{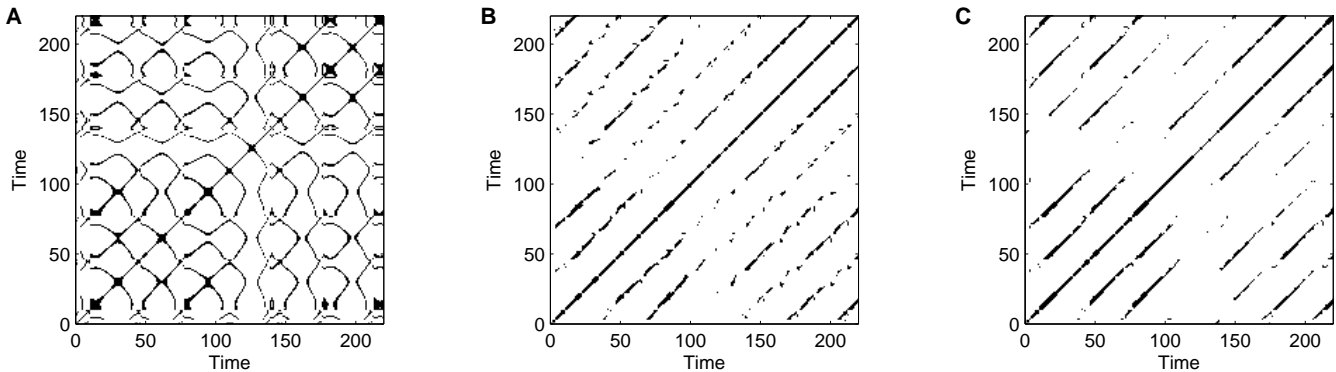} 
\caption{Recurrence plots of the R\"ossler oscillator 
with parameters $a=b=0.25$ and $c = 40$ using
different embedding:
(A) $m=1$, $\tau=1$, (B) $m=3$, $\tau=12$, (C) $m = 3$, $\tau = 6$
(adaptive recurrence threshold to ensure (A) $RR = 0.1$, (B, C) $RR = 0.05$).
Non-optimal embedding can cause line structures perpendicular to the main
diagonal, wobbly or interrupted lines (A, B).
}\label{rps_roessler_embeddings}
\end{figure}

If we start our recurrence analysis from a time series, we have first
to reconstruct a phase space by using a proper embedding, e.g.,
time-delay embedding \cite{packard80}. This involves the proper
setting of two additional parameters: the embedding dimension
$m$ and the time-delay $\tau$. Although the estimation of
dynamical invariants does not depend on the embedding \cite{thiel2004a},
the RQA measures depend on the embedding. Standard approaches
for finding optimal embedding parameters, like false 
nearest neighbours for embedding
dimension and auto-correlation or 
mutual information for time-delay, can be widely found
in the literature \cite[e.g.,][]{kantz97}. However, it is
recommended to visually cross-check the embedding parameters by 
looking at the resulting RP. Non-optimal embedding parameters 
can cause many interruptions of diagonal lines, small blocks,
or even diagonal lines perpendicular to the LOI (this corresponds
to parallel trajectory segments running in opposite time direction; 
Fig.~\ref{rps_roessler_embeddings}).
The experience has shown that the delay is sometimes overestimated
by auto-correlation and mutual information. The embedding dimension
has also to be considered with care, as it artificially increases
diagonal lines (will be discussed in Subsect.~\ref{sec_indic_det})
\cite{marwan2007}.

In general, it is recommended to
study the sensitivity (or robustness) of the results of 
the recurrence analysis
on the parameters (recurrence threshold, embedding parameters).

Although not really a parameter, it is worth to briefly discuss the 
different recurrence definitions. The most frequently used definition
is the to consider neighbours in the phase space which are smaller
than a threshold value (the recurrence threshold). Distances
can be calculated using different norms, like Maximum or Euclidean
norm \cite{marwan2007}. Maximum norm is sometimes preferred because
of its better computational efficiency (only minor differences in the
results when compared to Euclidean norm). Another definition of
recurrence considers a fixed amount nearest neighbours. This
recurrence criterion is used when the number of
neighbours is important. Pitfalls related to these recurrence
criteria are also discussed in Subsect.~\ref{sec_nonstat}.
More interesting are combinations of the above criteria with
dynamical properties of the phase space trajectory, e.g.,
perpendicular RPs (Subsect.~\ref{sec_indic_det}), or recurrence
based on order patterns \cite{groth2005}. Order patterns are
representations of the local rank order of a given number $d$ of 
values of the time series (order pattern dimension). 
As the number of order patterns is equal to $d!$, the dimension 
should not be chosen too large, because many order patterns
will appear rather seldom and the RP will be sparse. 
Even $d=4$ is often already not appropriate, therefore,
$d=3$ is the best choice in most cases (depending on the
problem of interest, $d=2$ may also be appropriate).

\subsection{Recurrence threshold selection}

%
%
%
%
%
%
%
%

The recurrence threshold $\varepsilon$ is a crucial parameter in the RP
analysis. Although several works have contributed to
this discussion \cite[e.g.,][]{thiel2002,matassini2002a,marwan2007,schinkel2008}, 
a general and systematic study on the recurrence threshold 
selection remains an open task for future work.
Nevertheless, recurrence threshold selection is a trade-off
of to have a small threshold as possible but at the same 
time a sufficient number of recurrences and recurrence
structures. 

However, the diversity of applicability of RP based methods
causes a number of different criteria for the selection
of the threshold: studying dynamical properties (dynamical
invariants, synchronisation) requires a very small threshold
\cite{marwan2007,donner2010a}; twin surrogates or trajectory reconstruction 
methods may require larger thresholds \cite{hirata2008};
noise corrupted observation data requires even larger 
thresholds \cite{thiel2002}; for studying dynamical transitions,
the threshold selection can be even without much importance, because 
the relative change of the RQA measures does not depend too much 
on it in a certain range; for the detection of certain signals
a specific fraction of the phase space diameter (or standard
deviation of the time series) can be required \cite{schinkel2008}.

Several ``rules of thumb'' for the choice 
of $\varepsilon$ have been advocated in the 
literature, e.g., a few per cent of the maximum phase 
space diameter \citep{mindlin92}, a value that
should not exceed 10\% of the mean or the maximum phase space diameter
\citep{koebbe92,zbilut92}, or that the recurrence rate $RR = \sum_{i,j} R_{i,j}/N^2$ 
is approximately 1\% \cite{zbilut2002b}. A recently proposed 
criterion employing the relationship between recurrence rate
and $\varepsilon$ defines an optimal value by using the 
position of the maximum of the first derivative of the 
recurrence rate $\frac{d RR}{d \varepsilon}$ \cite{gao2009}.
Such approach can produce ambiguous and highly unstable results,
as slight variations in $\varepsilon$ (as possible
by minor errors in finding this value or by nonstationary
time series) cause high variation in the recurrence structure.
Next, the position of the maximum of $\frac{d RR}{d \varepsilon}$ 
depends strongly on the chosen norm and embedding, and may lead
to an overestimation of an optimal $\varepsilon$. And,
finally, there are systems which can have more than one maximum 
\cite{donner2010a}.

Another criterion for the choice of $\varepsilon$ takes into account that a measurement
of a process is a composition of the real signal and some observational noise with 
standard deviation $\sigma$ \cite{thiel2002}. In order to get similar 
results as for the noise-free situation, $\varepsilon$
has to be chosen such that it is five times larger than the standard deviation 
of the observational noise, i.e.,~$\varepsilon > 5\,\sigma$.
Although this criterion holds for a wide class of processes, it is difficult
to estimate the amount of observational noise in the signal.

For (quasi-)periodic processes, it has been suggested to use 
the diagonal structures within the RP in order to find the
optimal $\varepsilon$ \cite{matassini2002a}. In this
approach, the density distribution of recurrence points along the 
diagonals parallel to the LOI is investigated on dependence of
$\varepsilon$ in order to minimise the fragmentation and thickness of 
the diagonal lines with respect to the threshold.
However, this choice of $\varepsilon$
may not preserve the important distribution of the diagonal lines in
the RP if observational noise is present (the estimated threshold
can be underestimated).

The selection of an optimal recurrence threshold $\varepsilon$
is not straightforward and depends on the particular problem and
question.

\subsection{Indicators of determinism}\label{sec_indic_det}
The length of a diagonal line in the RP corresponds to the time
the system evolves very similar as during another time, i.e., 
a segment of the phase space trajectory runs parallel and
within an $\varepsilon$-tube of another segment of the phase space 
trajectory. Deterministic systems are often characterised by
repeated similar state evolution (corresponding to a local predictability), 
yielding in a large
number of diagonal lines in the RP. In contrast, systems with independent
subsequent values, like white noise, have RPs with mostly single
points. Therefore, the fraction of recurrence points forming such 
diagonal lines (of length $l \ge l_{\min}$) 
\begin{equation}
DET = \frac{\sum_{l \ge l_{\min}} l P(l)}{\sum_{i,j} R_{i,j}}
\end{equation} 
can be calculated and is, therefore, called 
{\it determinism} in the RQA. Somehow this measure can be 
interpreted as an indication of determinism in the data.
But we should be careful in using the term determinism in
a more general or mathematical sense. In a deterministic system
we can calculate the same exact state by using given initial
conditions, i.e., there is no stochastic process involved.
Different methods can be used to test for determinism in
time series, e.g., a combined modelling-surrogate approach
\cite{small2003} or an analysis of the directionality 
of the phase space trajectory \cite{kaplan1992}.

\begin{figure}[bth]
\centering \includegraphics[width=\columnwidth]{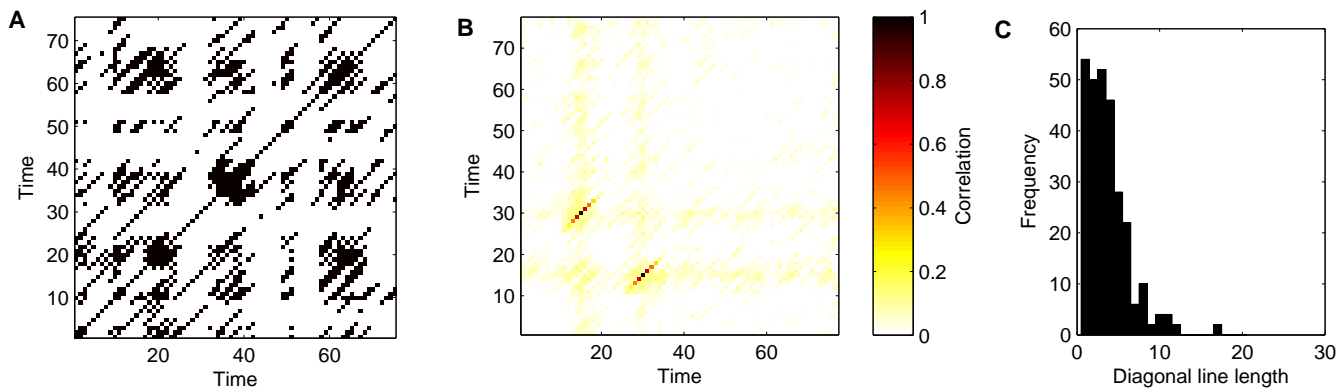} 
\caption{(A) Recurrence plot of one realisation of
Gaussian white noise, calculated
using embeddeding dimension $m=6$, delay $\tau=1$, and a recurrence threshold
of $\varepsilon=0.2$. The embedding causes a number of long lines.
(B) Correlation between a single recurrence point at $(15,30)$ and 
other recurrence points in the RP of white noise demonstrating
the effect of embedding for a bogusly creation of long 
diagonal lines (estimated from 1,000 realisations). 
(C) The histogram of line lengths found in the RP shown in (A). 
The maximum length is $L_{\max} = 17$, a value, which would not
be uncommon for a deterministic process.
}\label{rp_noise_embed}
\end{figure}

High values of $DET$
might be an indication of determinism in the studied system,
but it is just a necessary condition, not a sufficient one. 
Even for non-deterministic processes we can find longer
diagonal lines in the RP, resulting in increased $DET$ values.
For example, the following (non-deterministic) auto-regressive process 
$x_i = 0.8 x_{i-1} + 0.3 x_{i-2} - 0.25 x_{i-3} + 0.9 \xi$ (where 
$\xi$ is white Gaussian noise) has a $DET$ value of 0.6
(embedding dimension $m=4$, delay $\tau = 4$, and fixed recurrence
rate of 0.1). As it was shown in \cite{thiel2003},
stochastic processes can have RPs containing longer
diagonal lines just by chance (although very rare). Moreover,
due to embedding we introduce correlations in the RP and,
therefore, also uncorrelated data (e.g.~from white noise process)
have spurious diagonal lines \cite{thiel2006,marwan2007}
(Fig.~\ref{rp_noise_embed}). Moreover, data pre-processing
like low-passfiltering (smoothing) is frequently used.
Such pre-processing can also introduce spurious line structures
in the RP. Therefore, from just a high value of the RQA 
measure $DET$ we have to be careful in infering that the 
studied system would be deterministic. For such conclusion we
need at least one further criterion included in the
RP: the directionality of the trajectory \cite{kaplan1992}.
One possible solution is to use iso-directional
RPs \cite{horai2002} or perpendicular RPs \cite{choi99};
if then the measure reaches $DET \approx 1$ for
a very small recurrence density (i.e.~$RR<0.05$), the 
underlying system will be a deterministic one (like a periodic or chaotic 
system).

\subsection{Indicators of periodic systems}\label{sec_indic_period}
As explained in the previous section, deterministic systems
cause a high value in the RQA measure $DET$. This measure
has been successfully used to detect transitions in the
dynamics of complex systems \cite{trulla96}. A frequently
used example in order to present this ability is
the study of the different dynamical regimes
of the logistic map, where $DET$ is able to detect the periodic
windows (by values $DET = 1$). Therefore, it is often 
claimed that this measure is able to detect chaos-period 
transitions. 

\begin{figure}[bth]
\centering \includegraphics[width=.7\columnwidth]{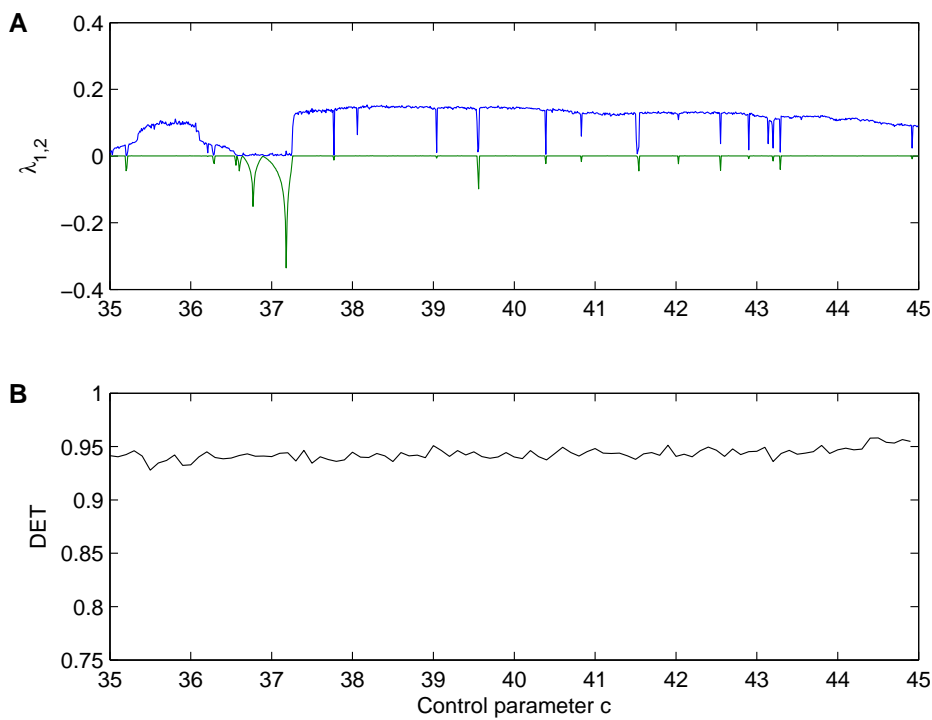} 
\caption{(A) 1st and 2nd positive Lyapunov exponents of the R\"ossler
oscillator with parameters $a=b=0.25$ and $c \in [35,\, 45]$. A periodic
window occurs between $c=36.56$ and $c=37.25$. However, the $DET$
measures reveals an almost constant very high value of approximately
$DET=0.94$. 
Used RP parameters: dimension $m=3$, delay $\tau=6$, 
adaptive recurrence threshold to ensure a $RR = 0.05$.
}\label{roessler_det}
\end{figure}

\begin{figure}[bth]
\centering \includegraphics[width=.5\columnwidth]{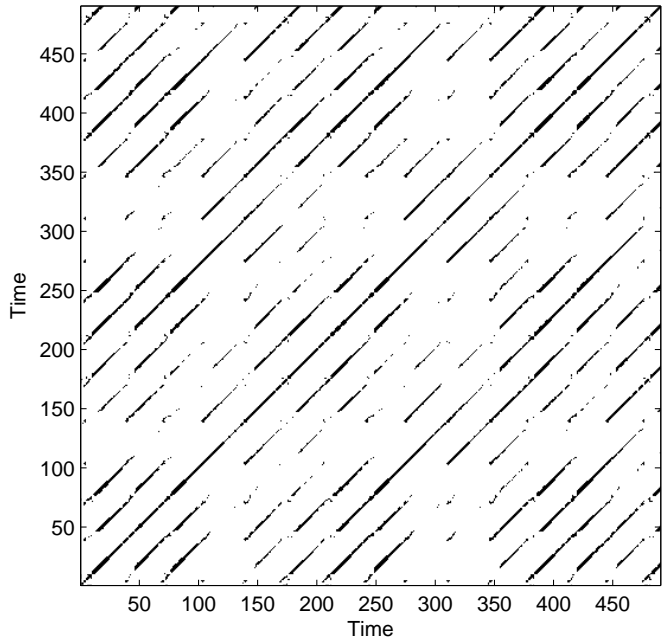} 
\caption{Recurrence plot of the R\"ossler
oscillator with parameters $a=b=0.25$ and $c = 40$. For this 
parameters, the R\"ossler system is in a chaotic regime ($\lambda_1 = 0.14$),
but the RP consists almost only on diagonal lines.
Used RP parameters: dimension $m=3$, delay $\tau=6$, 
adaptive recurrence threshold to ensure a $RR = 0.05$.
}\label{rp_roessler}
\end{figure}

However, we can also find
such high $DET$ values for non-periodic, but chaotic systems.
For example, the R\"ossler system \cite{roessler1976},
\begin{equation} \label{ros_eqs}
\left(\frac{dx}{dt},\,\frac{dy}{dt},\,\frac{dz}{dt}\right) =
\bigl(-y-z,\ x+0.25y,\ 0.25+z(x - c)\bigr),
\end{equation}
exhibits in the parameter interval $c \in [35,\,45]$ 
a transition from periodic to chaotic states (Fig.~\ref{roessler_det}A).
But due to the smooth phase space trajectory and high
sampling frequency (sampling time $\Delta t = 0.1$), the
RP for the chaotic trajectory consists almost exclusively 
on diagonal line structures (Fig.~\ref{rp_roessler}), 
resulting in a high value of 
$DET$, i.e., $DET \approx 1$ (Fig.~\ref{roessler_det}B).

A very high value of $DET$ is not a clear or even sufficient 
indication of a periodic
system. High values can be caused by very smooth phase space
trajectories. This should also be considered when looking
for indications of unstable periodic orbits (UPOs), where 
$DET$ or mean and maximal line lengths $L$ and $L_{\max}$
may not be sufficient. A solution could be to increase the minimal
length $l_{\min}$ of a diagonal recurrence structure which is 
considered to be a line. However, a better solution is to
look at the cumulative distribution of the diagonal line
lengths and estimate the $K_2$ entropy (but this requires
much longer time series, cf.~Subsect.~\ref{dyn_inv}).
Recent work has shown that measures coming from
complex network theory, like clustering coefficient,
applied to recurrence matrices are more powerful
and reliable for the detection of periodic dynamics 
\cite{marwan2009b,zou2010,donner2011}.

\subsection{Indicators of chaos}
The RP visualises the recurrence structure of the considered
system (based on the phase space trajectory). The basic idea
behind RPs comes, in general, from the study of chaos. Therefore it 
can be considered as a nonlinear tool for data analysis. 
But this cannot be a criterion to understand complex structures
in the RP or high values of RQA measures as indicators of 
chaos or nonlinearity in the dynamical system.

As mentioned above, uncorrelated stochastic systems have 
mostly short or almost no diagonal
line structures in their RPs, whereas deterministic and regular 
systems, like periodic processes, have mostly long and continuous diagonal 
line structures. Chaotic processes have also diagonal, but shorter 
lines, and can have single recurrence points. Nevertheless,
only by looking at the appearance of an RP it is difficult (almost
impossible) to infer about the type of dynamics; only periodic and
white noise processes can be identified with some certainty.  

The alternative is to look at the RQA measures quantifying the 
structures in an RP which are
related to some dynamical characteristics of the system.
As diagonal lines in the RP correspond to parallel running
trajectory segments, it is clear that the length of these
lines is somehow related to the divergence behaviour of the dynamical
system. Divergence rate of phase space trajectories is
measured by the Lyapunov exponent. In fact, the lengths of 
the diagonal lines are directly related to dynamical invariants as
$K_2$ entropy or $D_2$ correlation dimension 
\cite{faure98,thiel2004a}. The $K_2$ entropy is the lower limit
of the sum of the positive Lyapunov exponents.

For example, RQA measures based on the length of the 
diagonal lines, like determinism $DET$ and mean line length $L$, 
also depend on the type of the dynamics of the systems
(rather low values for uncorrelated stochastic 
(white noise) systems, higher values for more regular,
correlated and also chaotic systems). It has been suggested
to measure the length of the longest diagonal line $L_{\max}$
and interpret its inverse $DIV = 1/L_{\max}$ as an estimator
of the maximal Lyapunov exponent \cite{trulla96}. However,
this interpretation incorporates high potential of erroneous
conclusions derived from RQA. 

First, the main diagonal in the RP (i.e.,~the line of identity, LOI) 
is naturally the longest 
diagonal line, wherefore it is usually excluded from the analysis.
However, due to the tangential motion of the phase space 
trajectory\footnote{Tangential motion becomes even more
crucial and influential for highly sampled or smooth systems.},
subsequent phase space vectors are often also considered as 
recurrence points (known as sojourn points) \cite{marwan2007}.
These recurrence points lead to further continuous diagonal
lines directly close to the LOI. Without excluding an appropriate
corridor along the LOI (the Theiler window), $L_{\max}$ will be 
artificially large ($\approx N$) and $DIV$ too small. 

Second, as explained above, even white noise can have long
diagonal lines \cite{thiel2003}, leading to a small $DIV$
value just by chance (Fig.~\ref{rp_noise_embed}). 
Although the probability for the
occurrence of such long lines is rather small, the 
probability that lines of length two occur in RPs of stochastic 
processes is, on the contrary, rather high. Only one line of length two is
enough to get a finite value of $DIV$ which might be mis-interpreted
as a finite Lyapunov exponent and that the system would be chaotic
instead stochastic.

Therefore, we have to be careful in interpreting the RQA measures 
themselves as indicators of chaos. Moreover, 
such conclusion cannot be drawn by applying a simple
surrogate test where the data points are simply shuffled
(such a test would only destroy the correlation structure within 
the data, and, thus, the frequency information). 

RP or RQA alone cannot be used to infer nonlinearity from
a time series. For this purpose, advanced surrogate
techniques are more appropriate \cite{schreiber2000a,rapp2001}.


\subsection{Discrimination analysis and detection of deterministic signals}
RQA is also a powerful tool in order to distinguish
between different types of signals, different groups
of dynamical regimes etc.
\cite[e.g.,][]{zbilut98,marwan2002herz,facchini2007,litak2010}. 
However, the selection
of applicable RQA measures is a crucial task. Not
all measures will be useful for all questions. 
Their application needs justification in terms
of the purpose of the intended analysis. For example, for
processes which does not contain laminar regimes, 
or if we are not interested in the detection of
such laminar regimes, 
it would not make sense to use RQA measures basing
on vertical recurrence structures (like {\it laminarity}
or {\it trapping time}) \cite{marwan2009comment}.

\subsection{Indicators of nonstationarity and transition analysis}\label{sec_nonstat}
\begin{figure}[bth]
\centering \includegraphics[width=\columnwidth]{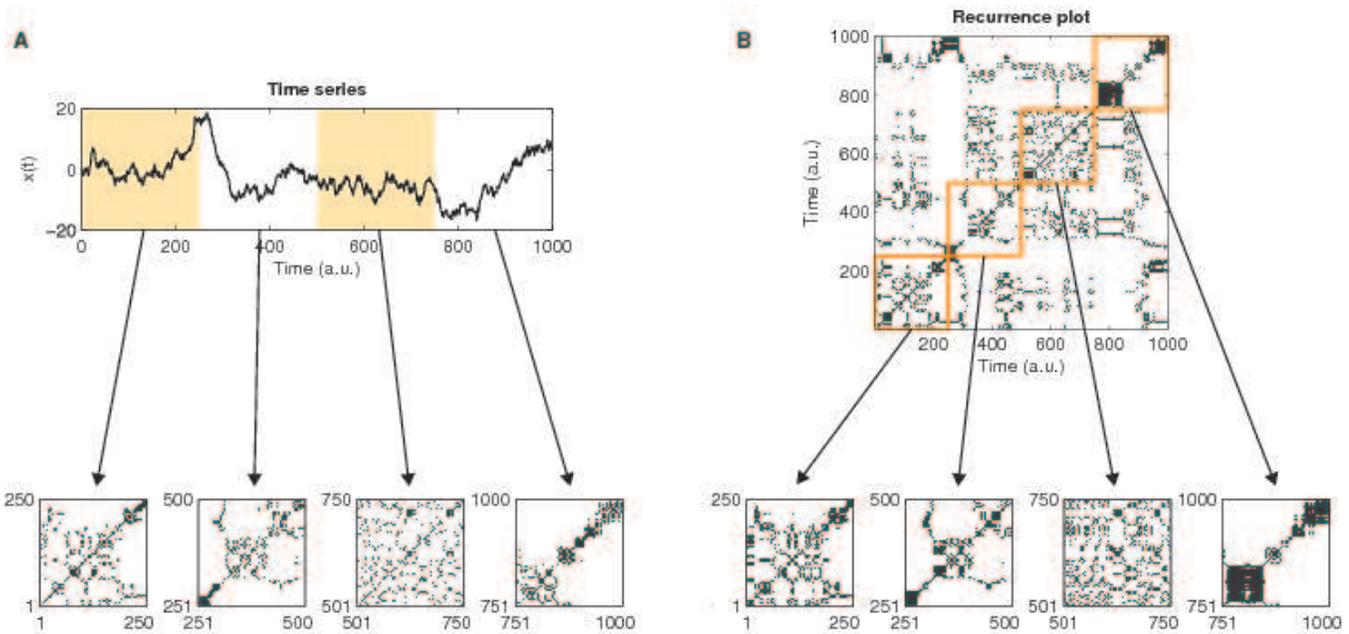} 
\caption{Two possibilities of windowed RQA: 
(A) Windowing of time series and
(B) windowing of RP.
The example is an auto-regressive process:
$x_i = 0.95 x_{i-1} + 0.05 x_{i-2} + 0.9 \xi$ (where 
$\xi$ is white Gaussian noise), the RP is calculated
using a constant number of neighbours (10\% of all
points) and without embedding. 
The sub-RPs at the bottom clearly demonstrate the
differences between the two approaches.
}\label{windowedRQA}
\end{figure}

RQA is powerful for the analysis of slight changes and transitions
in the dynamics of a complex system. For this purpose we need 
a time-dependend RQA (a RQA series) what can be realised in two ways
(Fig.~\ref{windowedRQA}):

(1) The RP is covered with small overlapping windows of size $w$ spreading
along the LOI and in which the RQA will be calculated, 
$R_{i,j}|_{i,j=k}^{k+w-1}$.

(2) The time series (or phase space trajectory) is divided into 
overlapping segments $x_i|_{i=k}^{k+w-1}$ from which RPs and 
subsequent RQA will be calculated separately.

%
%
%
%
%
%
%
%
%
%
%
%
%
%
%

Such time dependent approach can also be used to analyse the stationarity
of the dynamical system.

Here we should note the following important points. 
The time scale of the RQA values depends on the choice 
which point in the window should be considered as the
corresponding time point. Selecting the first point $k$
of the window as the time point of the RQA measures
allows to directly transfer the time scale of the 
time series to the RQA series. However, the window 
reaches into the future of the current time point
and, thus, the RQA measures represent a state which lies
in the future. Variations in the RQA measures can be
misinterpreted as early signs of later state transitions
(like a prediction). A better choice is therefore to select
the centre of the window as the current time point of
the RQA. Then the RQA considers states in the past and in
the future. If strict causality is required (crucial when
attempting to detect subtle changes in the dynamics just prior
the onset of dramatic state changes), it might
be even useful to select the end point of the window
as the current time point of the RQA (using embedding we have
to add $(m-1)\tau-1$). For most applications the
centre point should be appropriate.

Another important issue can rise from the different
windowing methods (1) or (2), which are only equivalent when we
do not normalise the time series (or its pieces) 
from which the RP is calculated and when we chose
a fixed threshold recurrence criterion. If we normalise
the time series just before the RP calculation, we
get differently normalised segments resulting in different
sub-RPs (and thus different RQA results) 
than such derived directly by moving windows from 
the RP of the entire time series (Fig.~\ref{windowedRQA} and
Tab.~\ref{tab_windowedRQA}). A similar problem 
arises when we use a fixed number of nearest neighbours
for the definition of recurrence, because it is a big
difference considering the entire time series in order
to find the $k$ nearest neighbours or just a small 
piece of it. Nevertheless, both approaches (1) and (2)
can be useful and depend on the given question. If we know
that the time series shows some nonstationarities or
trends which are not of interest, then approach (2)
can help to find transitions neglecting these nonstationarities.
But, if we are interested in the detection of the overall
changes (e.g., to test for nonstationarity), we should 
keep the numerical conditions for the
entire available time constant and chose approach (1). 
Anyway, for each RQA we should explicitly state how
the windowing procedure has been performed.

\begin{table}[tb]
\tbl{Selected RQA measures derived from windowing of
time series (left) and windowing of RP (right) of an
auto-regressive process and windowing 
as shown in Fig.~\ref{windowedRQA}.\label{tab_windowedRQA}}{
\begin{tabular}{lrrrr}
Window	&1--250		&251--500	&501--750	&751--1000\\
\hline
$RR$		&0.10		&0.10		&0.10		&0.10\\
$DET$		&0.62		&0.74		&0.48		&0.79\\
$L$		&3.13		&3.69		&2.75		&3.75\\
\end{tabular}
\quad
\begin{tabular}{lrrrr}
Window	&1--250		&251--500	&501--750	&751--1000\\
\hline
$RR$	&0.18		&0.12		&0.20		&0.19\\
$DET$		&0.81		&0.81		&0.69		&0.95\\
$L$		&3.78		&4.27		&2.90		&9.50\\
\end{tabular}
}
\end{table}

\begin{figure}[bth]
\centering \includegraphics[width=.5\columnwidth]{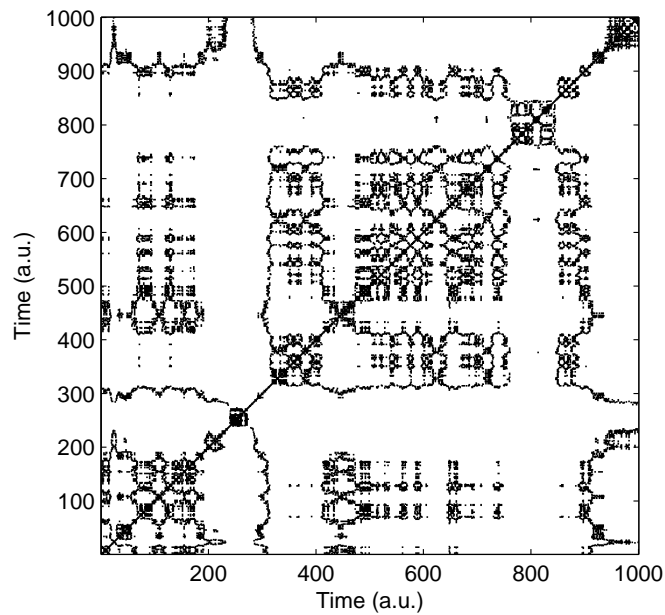} 
\caption{RP of the same auto-regressive process
as presented in Fig.~\ref{windowedRQA}, which is
by definition stationary. The RP is calculated
using maximum norm, $\varepsilon = 2$ and without embedding. 
}\label{ar_rp}
\end{figure}

\begin{figure}[bth]
\centering \includegraphics[width=.9\columnwidth]{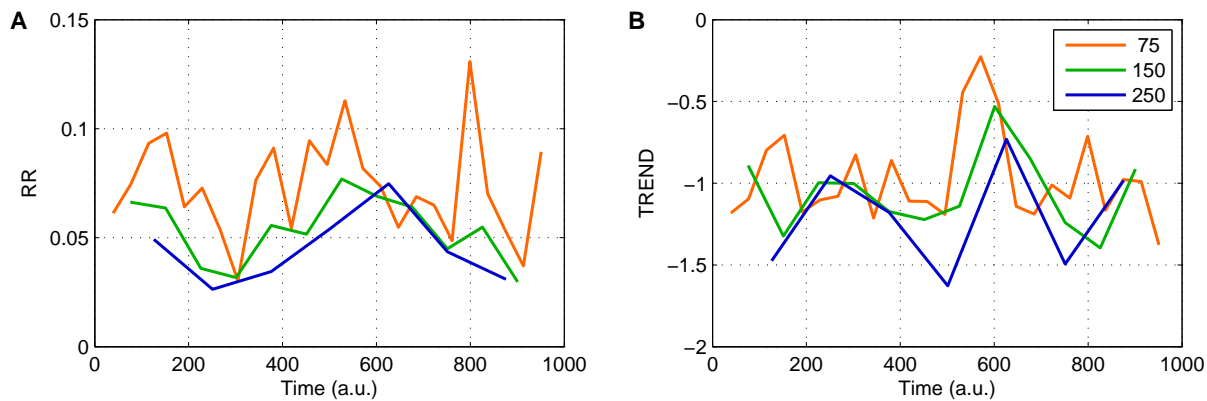} 
\caption{Two exemplary RQA measures, (A) recurrence
rate $RR$ and (B) paling trend ($TREND$), of the 
auto-regressive process as presented in Fig.~\ref{windowedRQA}
for three different window sizes $w$ ($w= 75, 150, 250$).
(A) The strong variation in $RR$ pretends a nonstationarity
in the signal. (B) $TREND$ depends rather strongly on $w$,
resulting in contrary outcomes, e.g., revealing
high values for $w=250$, but small values
for $w=75$ at the same time period $t=700 \ldots 800$.
The RQA is calculated 
using maximum norm, $\varepsilon = 0.3$ and without embedding
(the windows are moved by $w/2$, i.e., 50\% overlap; the
RQA time point is set to the centre of the RQA window).
}\label{nonstat}
\end{figure}

The choice of the window size itself needs the same
attention. Because the RQA measures are statistical measures
derived from histograms, the window should be large enough
to cover a sufficient number of recurrence lines or orbits.
A too small window can pretend strong fluctuations in
the RQA measures just by weak statistical significance
(the RQA measure $TREND$ is very sensitive to the 
window size and can reveal even contrary results, 
cp.~Fig.~\ref{nonstat}B). 
Therefore, conclusions about nonstationarity of the
system should be drawn with much care. Moreover, statements
on stationarity of the system itself are questionable at all
(if not enough knowledge about the system is available), 
because detected nonstationarity in an observed finite time series
does not mean automatically nonstationarity in the underlying 
system. For example, an auto-regressive process
is stationary by definition, but its RP and RQA can
reveal a nonstationary signal (Figs.~\ref{ar_rp} and \ref{nonstat}).

%
%
%
%
%
%
%
%

\subsection{Significance of RQA measures}

\begin{figure}[bth]
\centering \includegraphics[width=.9\columnwidth]{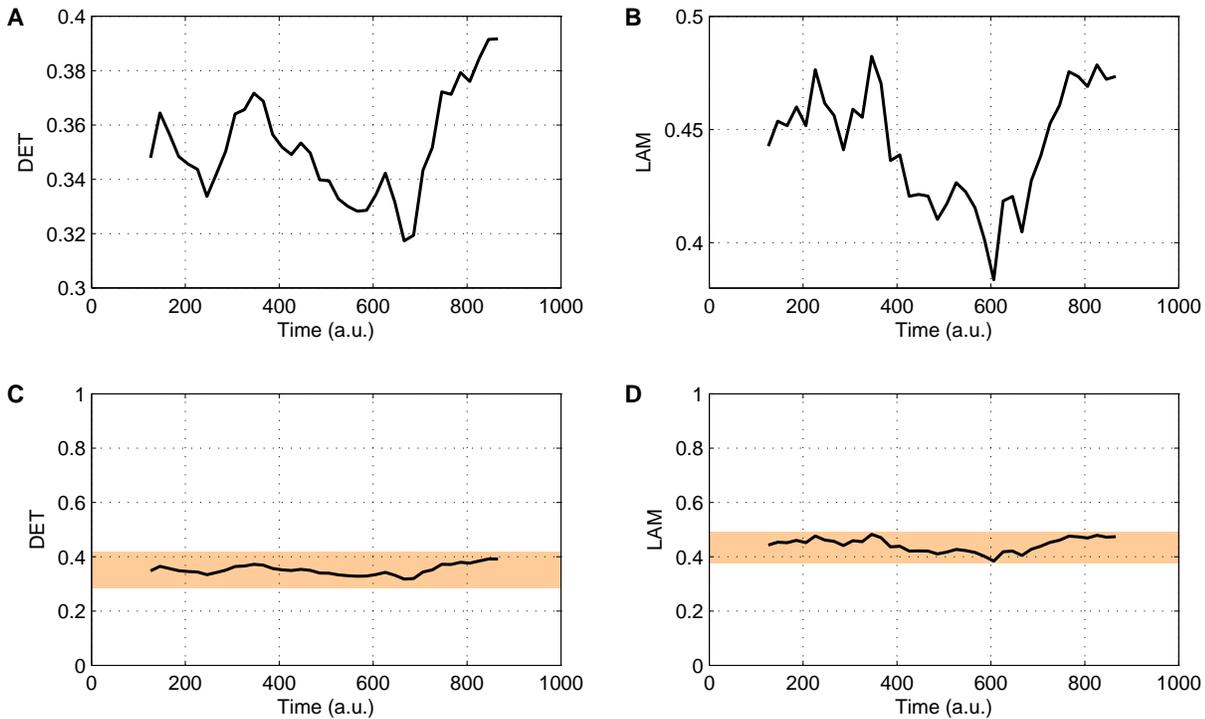} 
\caption{Two exemplary RQA measures, (A, C) determinism
$DET$ and (B, D) laminarity ($LAM$), of the 
auto-regressive process as presented in Fig.~\ref{windowedRQA}.
(A, B) The scaling of the $y$-axis is affecting a strong variation
in the RQA measures -- a potential of wrong conclusions.
(C, D) Considering a 5\% confidence interval of the RQA measures 
(details can be found in \cite{marwan2008nolta})
and a better value range for the $y$-axis, we cannot
infer that the values of the RQA measures as shown in (A) and (B)
significantly vary.
The RQA is calculated using a window size of $w=250$ and 
a window step of $ws = 20$, using maximum norm, 
$\varepsilon = 0.3$ and without embedding
(the
RQA time point is set to the centre of the RQA window).
$LAM$ is the fraction of recurrence points forming vertical
lines in an RP (analogously as $DET$ for the
diagonal lines).
}\label{signif}
\end{figure}

Related to the preceding issue on windowed RQA is the
question on the significance of the RQA variation.
A sub-optimal scaling of the variation of the RQA measures
can mislead to conclusions that the studied system
has changed its regime or that it would be nonstationary 
(Fig.~\ref{signif}A, B). Therefore, it is strongly recommended
to cross-check the scaling of the presentation and 
to present confidence intervals (Fig.~\ref{signif}C, D).
Confidence intervals can be calculated in various
ways, but we should avoid to derive them by simply
shuffling the original data. One approach could be
a bootstrap resampling of the line structures in the
RP \cite{marwan2008nolta,schinkel2009a}. Another approach
fits the probability of serial dependences (diagonal lines) 
to a binomial distribution \cite{hirata2011}. Whatever approach
we chose, the estimation of the confidence intervals is 
not a trivial task, but in the future the standard software
for RQA should include such tests.

%
%
%
%
%
%
%

A common statement on recurrence analysis is that it is useful
to analyse short data series. But we have to ask, how short is
short? The required length for the estimation of dynamical
invariants will be discussed in the following Subsect. Applying
RQA analysis we should be aware that the RQA measures are 
statistical measures (like an average) and need some minimal
length that a variation can be considered to be significant.

\subsection{Dynamical invariants from short time series}\label{dyn_inv}
An RP analysis is appropriate for analysing short and nonstationary
time series, as it is often stated in many reports 
\cite{fabretti2005,schinkel2007,zbilut98}. 
However, this statement holds actually only for the heuristic
measures of complexity as introduced for the RQA or for the
detection of differences or transitions in data series. If we are
interested in the dynamical invariants derived from RPs, the 
length $N$ of the time series becomes a more crucial part like 
it is for the standard methods of nonlinear data analysis. 

The derivations of dimensions ($D_1$, $D_2$) and 
dynamical invariants (like $K_2$) from
the RPs hold only in the limit $N \rightarrow \infty$ and small 
$\varepsilon$ ($\varepsilon \rightarrow 0$). Nevertheless,
an estimation of dynamical invariants from shorter time series 
can be feasible. We have to regard the following factors if discussing
the time series length: the number of orbits representing
stretching, the number of recurrences filling out a sufficient 
part of the attractor, and the number of data points necessary
for an acceptable phase space reconstruction \cite{wolf85}.
Since these factors may require different minimal lengths,
the largest of these lengths should be considered. 

For example, numerical considerations
for the estimation of the attractor (correlation) dimension
$D_2$ using the Grassberger-Procaccia algorithm 
\cite{grassberger83c} lead to the requirement 
$\log N > \frac{D_2}{2} \log(\frac{1}{\varrho})$ 
(where $\varrho = \frac{S}{\varepsilon}$ is the fraction the recurrence
neighbourhood of size $\varepsilon$ covers on the entire
phase space of diameter $S$) \cite{eckmann1992}. Considering a 
$\varrho = 0.1$ and a decimal logarithm, for finding a $D_2 = 10$ we
need at least $N=100,000$ data points. Furthermore, a $\varrho = 0.1$
is actually too large and we need much smaller $\varepsilon$,
which consequently provokes that again a larger $N$ is required.

For Lyapunov exponents (and analogously for $K_2$), a rough estimate 
based on the mentioned requirements 
suggests minimal time series lengths of $10^{D_2}$ to $30^{D_2}$ 
(with attractor dimension $D_2$) \cite{wolf85}. Accordingly, a 
system with $D_2=3$ requires 1000--30,000 data points (a more strict
consideration even requires 
$\log N > D_2 \log(\frac{1}{\varrho})$ \cite{eckmann1992}).

Therefore, too guarantee useful results we need long time series.
If we calculate dimensions or $K_2$ from short time series the results
are probably worthless.

\subsection{Synchronisation and line of synchronisation}
Cross recurrence plots (CRPs) can be used for the investigation of 
the simultaneous evolution of two different phase space trajectories
\cite{marwan2002npg,marwan2002pla,zolotova2009,ihrke2009}.
The {\it line of identity (LOI)} in the RP becomes a {\it line
of synchronisation (LOS)} in the CRP. Two more-or-less identical
systems but with differences on the time-scale will reveal
a bowed LOS \cite{marwan2002npg,marwan2005}. An off-set of the LOS
away from the main diagonal is an indication of a phase shift
or a delay between the two considered systems. 

However, because this method tests if the two trajectories
visit the same region in the phase space, it can be used only 
to study complete synchronisation (CS) or a kind of a generalised
correlation (although with possible delays), or to get the 
relation between the transformations between their time-scales.
Moreover, the data under consideration should be from the
same (or a very comparable) process and, actually, should represent the same 
observable. Therefore, the reconstructed phase space should be the same.

For the study of the LOS the distance matrix may
be more appropriate because it contains more information,
especially if the data series show nonstationarities. Then,
the LOS can be found by using efficient algorithms like
dynamic time warping \cite{sakoe1978}. Nevertheless,
it is always very important to check if the found LOS 
makes sense; for instance, it is possible to find
several LOS
(cp.~Application in magneto-stratigraphy in \cite{marwan2007}).

\subsection{Macrostructures and sampling}

\begin{figure}[hbtp]
\centering \includegraphics[width=.9\columnwidth]{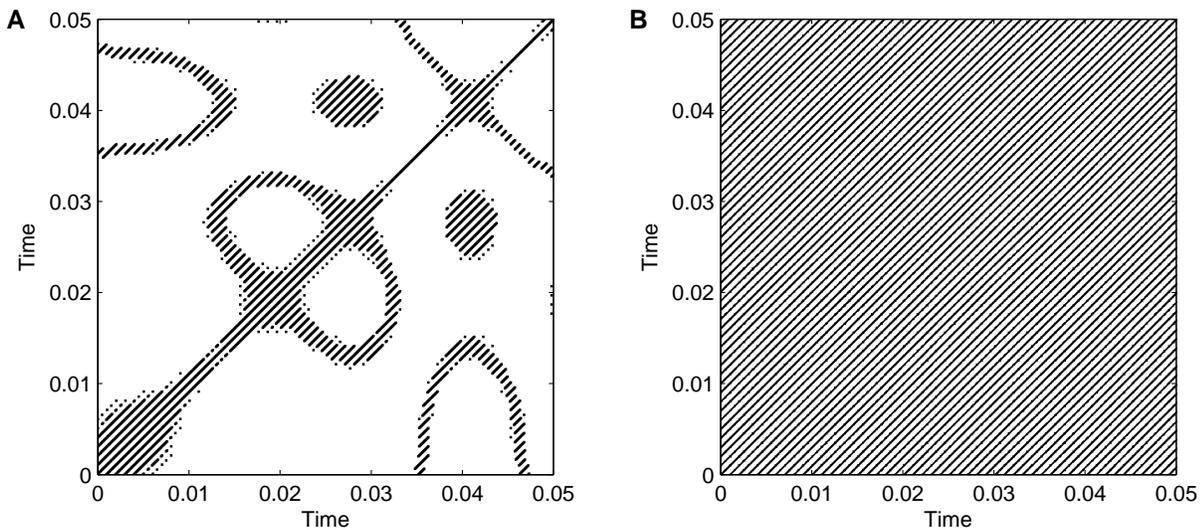}
\caption{RP of a modulated harmonic oscillation 
$\sin(2 \pi 1000(\pi+t) + 2 \pi \sin(2\pi 44 t)t)$.
(A) Non-trivial macrostructures (gaps) in the RP
due to the interference of the sampling frequency of 1~kHz and
the frequency of the modulated harmonic signal.
(B) Corresponding RP as shown in (A), but for a higher sampling frequency
of 10~kHz. As expected, the entire RP now consists of the
periodic line structures due to the oscillation.
Used RP parameters: dimension $m=3$, delay $\tau=1$, 
recurrence threshold $\varepsilon=0.05\sigma$, 
$L_{\infty}$-norm.
}
\label{macrostruct2}
\end{figure}

\begin{figure}[htb]
\centering \includegraphics[width=.9\columnwidth]{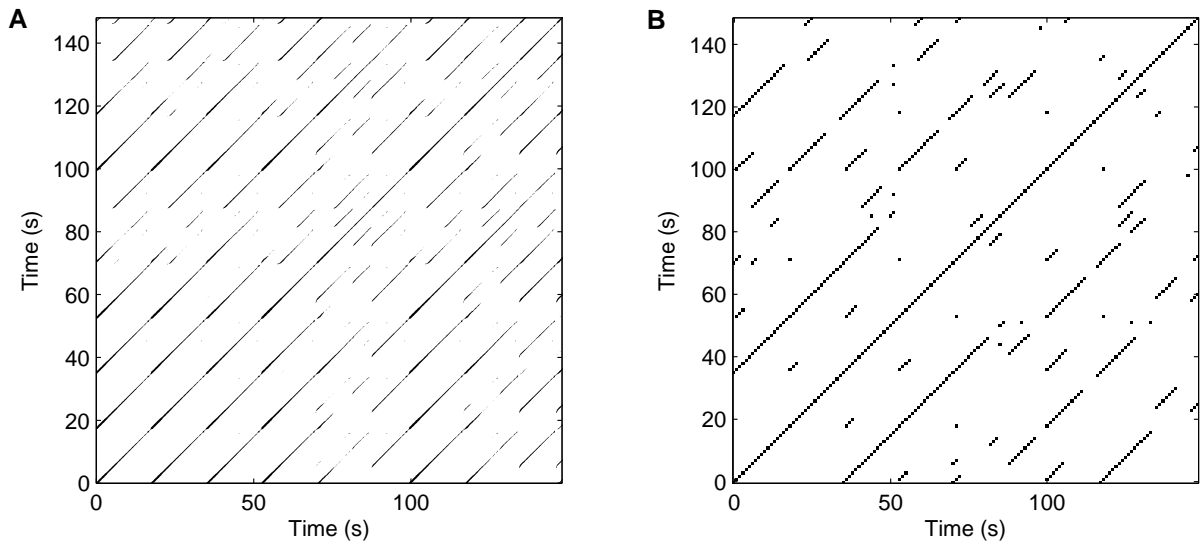} 
\caption{RP of the $x$-component of the R\"ossler oscillator, 
Eq.~(\ref{ros_eqs}), with parameters $a=b=0.2, c=5.7$. The 
sampling time is (A) $\Delta t = 0.05$ and (B) $\Delta t = 1$.
The embedding was chosen in both settings to be equivalent:
dimension in (A) and (B) is $m=3$, the  delay in 
(A) $\tau = 20$, but in (B) $\tau = 1$; recurrence threshold
$\varepsilon = 1.5$ (maximum norm).
Due to the low sampling in (B), many diagonal lines vanish.
}\label{macrostruct}
\end{figure}

For the visual interpretation of an RP and also for a reliable RQA
we should remember that our data are discretised time or data series.
The sampling of the signal has an importance which should not be underestimated.
If the sampling frequency
is just one magnitude higher than the system's main frequencies, and their ratio
is not a multiple of an integer (i.e., we have an intrinsic phase error), 
an interference triggered by the sampling of the continuous signal can
produce large empty regions in the recurrence matrix, although
they should be there \cite{facchini2005,facchini2007a}. Nonstationarities
or modulations in frequency or phase cause non-trivial gaps or
macrostructures in the recurrence matrix (Fig.~\ref{macrostruct2}). 
We should be aware that such gaps can occur in particular when we use
a low sampling frequency. The recurrence structure of interest can appear
rather different; diagonal lines can vanish or can be reduced to just single
points yielding in biased RQA measures (Fig.~\ref{macrostruct}).

%
%

Nevertheless, tiny modulations in frequency or phase in oscillating signals can 
be detected by RPs, which are non-detectable by standard methods
(spectral or wavelet analysis). This turns RPs to a powerful tool
for the analysis of slight modulations in oscillatory signals
like audio signals.

\begin{figure}[bth]
\centering \includegraphics[width=.7\columnwidth]{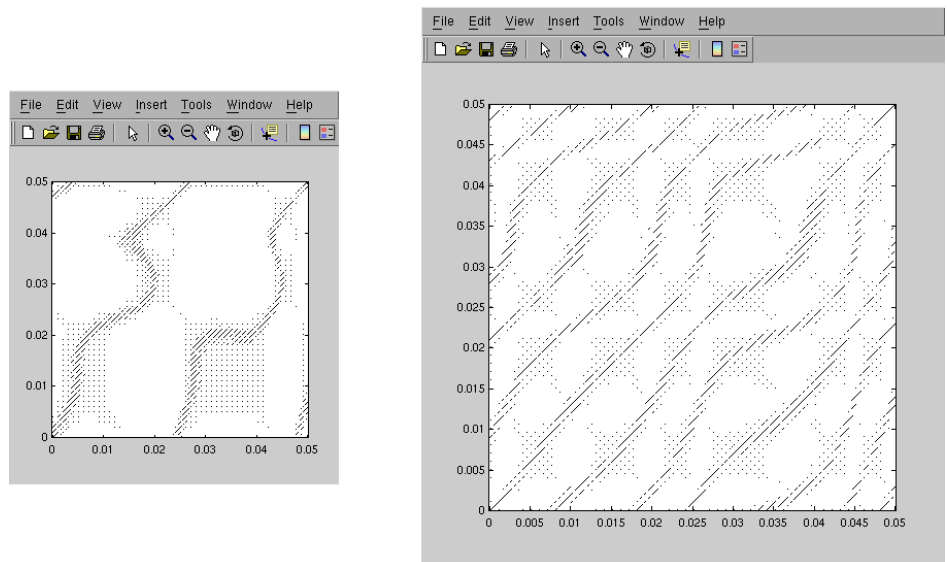} 
\caption{Screenshot of the RP as shown in
Fig.~\ref{macrostruct2}B for two different window sizes
of display on a computer screen (using 
Matlab\textsuperscript{\textregistered}). Although the
RP consists only on continuous diagonal lines as represented
in Fig.~\ref{macrostruct2}B, its size ($N = 5511$) exceeds
the screen resolution and requires downsampling, leading
to artifical macrostructures.
}\label{rp_screen}
\end{figure}

Please note that macrostructures are also an apparent problem
when displaying large RPs on a computer screen (and up to
a certain amount on print outs). The resolution of
modern computer screens is around 72~ppi (points per inch,
72~ppi corresponds to around 28 points per
centimetre). The presentation of RPs in a window of, e.g., 6~inch
allows only the display of around 430 points. Larger RPs will
be rendered using downsampling or interpolation, resulting in 
similar interference effects and artificial secondary macrostructures
as described above; such macrostructures will even change for
different window sizes (Fig.~\ref{rp_screen}). 
Therefore, we should take care in 
visual interpretation of patterns found in large RPs which are
represented on computer screens.

%
%
%
%
%

%

\section{Conclusions}
We have illustrated several problems regarding the application
of recurrence plots (RPs) and recurrence quantification analysis
(RQA) which need our attention in order to avoid wrong
results. The uncritical application of these methods can
yield to serious pitfalls. Therefore, it is important to
understand the basic principles and ideas behind the measures
of complexity forming the RQA and the different techniques
to study the numerous phenomena of complex systems, like
transitions, synchronisation, etc. Nevertheless, the
recurrence plot based techniques are still a rather young
field in nonlinear time series analysis, and many open
questions remain. For example, systematic research
is necessary to define reliable criteria for the selection
of the recurrence threshold, and the estimation of the 
confidence of the RQA measures will be a hot topic in the
near future.

\nonumsection{Acknowledgments} 
\noindent The work has been supported by the Potsdam Research Cluster for
Georisk Analysis, Environmental Change and Sustainability
(PROGRESS).

\bibliographystyle{ws-ijbc}
\bibliography{rp,mybibs}

\end{document}